\documentclass[aps,prd,a4paper,superscriptaddress,nofootinbib,10pt, two column]{revtex4-1}
\usepackage{amsmath,amssymb,graphicx,multirow,dcolumn,bm,latexsym,soul,nicefrac}
\usepackage[colorlinks,linkcolor=blue,citecolor=blue,urlcolor=blue]{hyperref}
\newcounter{RomanNumber}
\usepackage[normalem]{ulem}

\allowdisplaybreaks
\newcommand{\be}{\begin{equation}}
\newcommand{\ee}{\end{equation}}
\newcommand{\bea}{\begin{eqnarray}}
\newcommand{\eea}{\end{eqnarray}}
\newcommand{\nn}{\nonumber}

\def\simlt{\mathrel{\lower0.6ex\hbox{$\buildrel {\textstyle <}
 \over {\scriptstyle \sim}$}}}

\def\bea{\begin{eqnarray}}
\def\eea{\end{eqnarray}}
\def\be{\begin{equation}}
\def\ee{\end{equation}}
\def\bes{\begin{split}}
\def\ees{\end{split}}
\def\ba{\begin{eqnarray}}
\def\ea{\end{eqnarray}}
\def\p{\partial}

\def\dd{{\rm d}}
\def\dt{{\rm d}t}
\def\dv{{\rm d}v}
\def\dr{{\rm d}r}
\def\dx{{\rm d}x}

\def\dz{{\rm d}z}

\def\ds{{\rm d}s}

\def\dOm{{\rm d}\Omega}

\def\m{{\rm m}}
\def\({\Big(}
\def\){\Big)}

\def\mO{{\mathcal O}}

\def\a{\alpha}
\def\b{\beta}
\def\c{\chi}
\def\r{\rho}
\def\s{\sigma}

\def\m{\mu}
\def\n{\nu}

\def\S{\Sigma}

\def\l{\lambda}

\def\d{\delta}
\def\kh{{\rm kh}}

\newcommand{\lmn}{_{\mu\nu}}

\def\vv{{\rm v}}
\def\dv{{\rm dv}}

\def\mn{{\mu\nu}}

\begin{document}

\title{Neutron star sensitivities in Ho\v rava gravity after GW170817}

\author{Enrico Barausse}
\affiliation{SISSA, Via Bonomea 265, 34136 Trieste, Italy and INFN Sezione di Trieste}
\affiliation{IFPU - Institute for Fundamental Physics of the Universe, Via Beirut 2, 34014 Trieste, Italy}
\affiliation{Institut d'Astrophysique de Paris, CNRS \& Sorbonne
 Universit\'es, UMR 7095, 98 bis bd Arago, 75014 Paris, France}

\begin{abstract}
Ho\v rava gravity breaks boost invariance in the gravitational sector by introducing
a preferred time foliation. The dynamics of this preferred slicing is governed, in the low-energy limit
suitable for most astrophysical applications, by three dimensionless parameters $\alpha$, $\beta$ and $\lambda$.
The first two of these parameters are tightly bound 
by solar system and gravitational wave propagation experiments, but $\lambda$ remains relatively unconstrained ($0\leq\lambda\lesssim 0.01-0.1$).
We restrict here to the parameter space region defined by $\alpha=\beta=0$ (with $\lambda$ kept generic), 
which in a previous paper we showed to be the only one where black hole solutions are non-pathological at the universal horizon, and
 we focus on possible violations of the strong equivalence principle in systems involving neutron stars. 
We compute neutron star ``sensitivities'', which parametrize violations of the strong equivalence principle at the leading post-Newtonian order,
and find that they vanish identically, like in the black hole case, for $\a=\b=0$ and generic $\l\neq0$. This implies
that no violations of the strong equivalence principle (neither in the conservative sector nor in gravitational wave fluxes) 
can occur at the leading post-Newtonian order in binaries of compact objects, and that data from binary pulsars and gravitational interferometers are unlikely to 
further constrain $\l$.
\end{abstract}
\maketitle

\section{Introduction}\label{intro}
Since Lorentz invariance is widely regarded as a fundamental symmetry of Nature, 
much effort has been devoted to test it experimentally, 
 both in the pure matter~\cite{Kostelecky:2003fs,Kostelecky:2008ts,Mattingly:2005re,Jacobson:2005bg} and pure gravity~\cite{Jacobson:2008aj,Liberati:2013xla} sectors, and in
matter-gravity interactions~\cite{Kostelecky:2010ze}. However, 
bounds on this symmetry are still considerably weaker in 
gravitational experiments than in particle physics ones. This is 
due both to the intrinsic weakness of the gravitational interaction (with the 
resulting difficulties of performing suitable experiments, which
need to be very accurate and/or rely on astrophysical systems)
and to the absence of a generic parametrization of all possible Lorentz-violating effects in the strong 
gravity, highly relativistic regime. (See instead
e.g. Refs.~\cite{Colladay:1998fq, Kostelecky:1998id,Kostelecky:1999rh}
 for a Standard Model extension suitable for describing parametrized Lorentz violations in matter.)

One possible approach is to consider the simplest gravitational theory that extends general relativity and which
breaks Lorentz symmetry at low energies in  a dynamical fashion. Focusing in particular on boost symmetry, 
the simplest way to break it is to introduce a 
preferred time direction at each spacetime event, by means of a timelike, unit-norm vector field (the ``\ae ther''). 
To make this vector field dynamical, the gravitational action must include a kinetic term, whose form
is fixed (apart from dimensionless coupling constants) if we require that it should be covariant and quadratic in the \ae ther derivatives~\cite{Jacobson:2000xp}.
Moreover, one may require that the \ae ther should define not only a preferred time direction locally, but also a preferred
time foliation of the spacetime manifold. To this purpose, the \ae ther vector needs to be  hypersurface orthogonal, i.e.
proportional to the gradient of a timelike scalar (which we will refer to as the ``khronon''~\cite{Blas:2009qj}),
whose level sets define the preferred time foliation. 
The theory obtained in this way, known as ``khronometric theory''~\cite{Horava:2009uw,Blas:2009qj,Jacobson:2010mx}, is the simplest theory
that breaks boost invariance at low energies. A related theory, referred to as ``Einstein-\ae ther theory''~\cite{Jacobson:2000xp}, is obtained 
by waiving the requirement that the \ae ther be hypersurface orthogonal.

In the following we will focus solely on khronometric theory. Indeed, this theory is not only interesting
from a phenomenological point of view as a framework to test Lorentz invariance in the gravitational sector, but it
also arises naturally as the low-energy limit of Ho\v rava gravity~\cite{Horava:2009uw}. The latter is a gravitational
theory that, unlike general relativity, is power-counting and perturbatively renormalizable~\cite{Horava:2009uw,Barvinsky:2015kil}. As such, it
provides a possible model for quantum gravity, although it remains to be seen whether
the percolation of Lorentz violations from the gravitational sector to the matter one can
be sufficiently suppressed to satisfy the tights bounds on Lorentz violations in matter. 
Note that this percolation may be suppressed via a large energy scale~\cite{Pospelov:2010mp}, or
Lorentz invariance may emerge in the  infrared
as a result of renormalization group flows~\cite{Chadha:1982qq,Bednik:2013nxa,Barvinsky:2017kob} (see however also Ref.~\cite{Knorr:2018fdu})
 or  accidental symmetries~\cite{GrootNibbelink:2004za}.

Bounds on khronometric theory can be obtained in a number of ways. Theoretical
requirements such as the absence of classical (gradient)  and quantum (ghost) instabilities~\cite{Blas:2011zd,Jacobson:2004ts,Garfinkle:2011iw}, combined
with solar system bounds on the preferred frame parameters~\cite{Will:2014kxa,Blas:2011zd,Bonetti:2015oda} and the absence of vacuum Cherenkov radiation from ultrahigh energy cosmic rays~\cite{Elliott:2005va}, have been used to constrain the dimensionless coupling parameters ($\a,\,\b$ and $\l$) of the theory. 
Moreover, Refs.~\cite{Yagi:2013ava,Yagi:2013qpa} considered the evolution of binary pulsars under gravitational wave emission, 
and showed that timing of these systems, when combined with cosmological
observations (and namely measurements of the 
abundance of primordial elements produced during Big Bang Nucleosynthesis~\cite{Carroll:2004ai})  allowed
for bounding the coupling constants to within a few percent.

More recently, the coincident gravitational wave and electromagnetic detection of GW170817 and GRB 170817A~\cite{GBM:2017lvd,PhysRevLett.119.161101}
has constrained the propagation speeds of gravitational waves and light to match to within one part in $10^{15}$~\cite{Monitor:2017mdv}. This bounds
one of the parameters of the theory, $\b$, to $|\b|\lesssim 10^{-15}$~\cite{Gumrukcuoglu:2017ijh,ramos}. Further bounds follow from
combining the  GW170817/GRB 170817A observations with solar system constraints
on the preferred frame parameters~\cite{Gumrukcuoglu:2017ijh,ramos}. Indeed, if one imposes $|\b|\lesssim 10^{-15}$, the latter are satisfied by $|\a|\simlt 10^{-7}$, at least if $|\l|\gg 10^{-7}$;
or by $|\a|\simlt 0.25\times 10^{-4}$ \textit{and} $\lambda\approx\alpha/(1-2 \alpha)$~\cite{ramos}.
This second possibility implies small values for all three coupling constants ($|\alpha|\sim |\lambda|\lesssim 10^{-5}$ and $|\b|\lesssim 10^{-15}$),
while the first tightly constrains  $\alpha$ and $\beta$ ($|\a|\simlt 10^{-7}$, $|\b|\simlt 10^{-15}$) but leaves $\lambda$ essentially unconstrained.
The only bounds on $\l$ are then $\l\lesssim 0.01-0.1$, from measurements of the abundance of primordial elements produced during Big Bang Nucleosynthesis~\cite{Carroll:2004ai,Yagi:2013ava,Yagi:2013qpa}  and from
 Cosmic Microwave Background observations~\cite{Afshordi:2009tt}; and the sign requirement $\l\geq0$
needed to ensure the absence of ghosts~\cite{Blas:2009qj,Gumrukcuoglu:2017ijh}.
In the light of these constraints, it is therefore quite tempting to simply set $\a$ and $\b$ exactly to zero while keeping
$\l$ finite.

Ref.~\cite{ramos} has recently studied the structure of black holes moving slowly relative to the preferred foliation
in khronometric theory. Indeed, those systems are key to understanding gravitational wave emission from black hole binaries
at the lowest post-Newtonian (PN) order, and to computing the ``sensitivities''~\cite{Foster:2007gr,Yagi:2013ava,Yagi:2013qpa,1975ApJ...196L..59E,0264-9381-9-9-015,Will:1989sk,Barausse:2015wia,Yagi:2015oca,ramos} that parametrize dipole gravitational emission and 
violations of the strong equivalence principle (i.e. violations of the universality of free fall for
strongly gravitating objects). A surprising finding of Ref.~\cite{ramos} is that these 
slowly moving black holes present an unavoidable finite-area curvature singularity at the ``universal horizon''~\cite{Barausse:2011pu,Blas:2011ni} (i.e.
at the causal boundary for signals propagating with arbitrarily large speeds in the ultraviolet limit, which exist in Ho\v rava gravity as a result of
Lorentz violations), unless the parameters $\a$ and $\b$ are set exactly to zero. Furthermore, 
if one sets $\a=\b=0$ while keeping $\l\neq0$, the geometry of these black holes is exactly the same as in general relativity
and their sensitivities vanish, even though the khronon profile is non-trivial~\cite{ramos}. This entails in particular that gravitational wave emission from binary black holes matches general relativity
at the leading PN order, and no vacuum gravitational dipole radiation exists for $\a=\b=0$.

Once ascertained that it makes sense to set $\a$ and $\b$ exactly to zero to satisfy the experimental bounds presented above and to
ensure that curvature singularities do not appear in moving black holes, an obvious question is how gravitational wave emission in khronometric
theory behaves in systems involving neutron stars, and more in general whether such systems
can present violations of the strong equivalence principle. Of course, this is relevant to compare both to pulsar timing observations~\cite{Yagi:2013ava,Yagi:2013qpa}
and to direct gravitational detections of binary neutrons stars, such as GW170817~\cite{GBM:2017lvd,PhysRevLett.119.161101}.\footnote{Note that the aforementioned bounds on $\b$ from GW170817 and GRB 170817A
follow from the propagation of the gravitational wave signal alone, but do not exploit its generation.} This problem was studied in detail
in Refs.~\cite{Yagi:2013ava,Yagi:2013qpa}, as we have mentioned above, but the parameter space considered in those works does not include the case 
$\a=\b=0$ now favored by both the multimessenger detection of GW170817 and GRB 170817A~\cite{Monitor:2017mdv,Gumrukcuoglu:2017ijh,ramos}, and by the results of Ref.~\cite{ramos} on the regularity of
moving black holes. 

Here, we will fill this gap, and show that for $\a=\b=0$ stellar sensitivities vanish exactly, which in turn implies
that the dynamics of binary neutron stars matches general relativity in both the conservative and dissipative sectors at leading order in the PN expansion.
Therefore,
in particular, no gravitational dipole radiation is emitted in khronometric gravity in the viable region of parameter space (unlike what happens
for generic $\a$ and $\b$, c.f. Refs.~\cite{Yagi:2013ava,Yagi:2013qpa}), 
and emission at quadrupole order, as well as the conservative Newtonian dynamics, matches the general relativistic prediction. As a result, binary pulsars/GW170817-like observations
are unlikely to be useful to constrain the remaining theory parameter $\l$.

This paper is organized as follows. In Sec.~\ref{horava} we will briefly review the action and field equations of khronometric theory; in Sec.~\ref{GW} we will review
 the concept of compact object sensitivities and their effect on the dissipative and conservative dynamics; in Sec.~\ref{sec_ans} we will derive the ans{\"a}tze for the metric and khronon of stars moving slowly relative to the preferred frame; in Sec.~\ref{ext} we will
solve the field equations for these stars for $\a=\b=0$,
and show that their geometry matches exactly the predictions of general relativity (implying in particular that the sensitivities vanish exactly).
Finally, in Sec.~\ref{disc} we will further discuss our results and put forward our conclusions.
Throughout this paper, we will set $c=1$ and use a metric signature $(+,-,-,-)$.

\section{Khronometric theory}\label{horava}
The low energy limit of Ho\v rava gravity, which is suitable for most astrophysical applications involving compact objects~\cite{Barausse:2011pu,Barausse:2013nwa}, can be described covariantly by the khronometric theory action~\cite{Jacobson:2010mx}
\begin{multline}\label{khaction}
S=-\frac{1}{16\pi G} \int \dd^4x\, \sqrt{-g} \Big[R+\l\;(\nabla_\m u^\m)^2\\+\b\nabla_\m u^\n \nabla_\n u^\m+\a\; a_\m a^\m\Big]+S_{\rm matter}[g_{\mu\nu},\Psi]\,,
\end{multline}
where $\a,\,\b$ and $\l$ are dimensionless coupling constants, $a^\m\equiv u^\n \nabla_\n u^\mu$, and
 the timelike unit-norm ``\ae ther'' vector field $\boldsymbol{u}$ is given by
\be \label{khaether}
u_\mu=\frac{\nabla_\mu T}{\sqrt{\nabla^\alpha T\nabla_\alpha T}}\,,
\ee
with $T$ a timelike scalar field (the ``khronon'') defining the preferred time foliation.
The matter fields $\Psi$ are coupled minimally to the four-metric $g_{\mu\nu}$ alone, so as to enforce geodesic motion for matter and the weak equivalence principle at tree level. This minimal coupling implies in particular that the bare gravitational constant $G$ must be related to the locally
measured $G_N$ by $G_{N}={G}/({1-\alpha/2})$~\cite{Carroll:2004ai}. 

Note that the \ae ther vector can be ``eliminated'' from this action by choosing coordinates adapted to the khronon, i.e. by choosing 
the khronon as the time coordinate. This yields $u_\mu=N \delta_\mu^T$ (where $N$ is the lapse), while the action takes the non-covariant form~\cite{Horava:2009uw,Blas:2009qj,Blas:2010hb}
\be
\begin{split}
S=&\frac{1-\beta}{16\pi G}\int \dd T\dd^3x\, N\sqrt{\gamma} \Big(K_{ij}\,K^{ij}-\frac{1+\lambda}{1-\beta}K^2\\
&+\frac{1}{1-\beta}{}^{(3)}{R}+\frac{\alpha}{1-\beta}a_i\,a^i\Big)+S_{\rm matter}[g_{\mu\nu},\Psi]\,,
\end{split}
\ee 
where $K^{ij}$, ${}^{(3)}{R}$ and $\gamma_{ij}$ are  the extrinsic curvature, three-dimensional Ricci scalar and three-metric of the
$T=$ const foliation; $K=K^{ij}\gamma_{ij}$; and $a_i\equiv\p_i\ln N$. This action, which is
invariant under the foliation-preserving diffeomorphisms
\be
T\rightarrow \tilde{T}(T)\,,\qquad x^i\rightarrow \tilde{x}^i(x,T)\,,
\ee
but {\it not} under full four-dimensional diffeomorphisms, is the infrared limit of the original  action proposed by Ho\v rava~\cite{Horava:2009uw}. Nevertheless, in the following
we will find it more convenient to use the equivalent covariant action \eqref{khaction}.

By varying the action (\ref{khaction})
with respect to $g^{\mu\nu}$ while keeping $T$ fixed, one gets the modified Einstein field equations~\cite{Barausse:2012qh,Barausse:2013nwa}
\begin{equation}\label{khEinsteinEq}
        G_\mn- T^{\kh}_\mn=8\pi G \, T_\mn^{\rm matter}\,,
\end{equation}
where $G\lmn=R\lmn-R\, g\lmn/2$ is the Einstein tensor, 
$T^\mn_{\rm matter}=({-2}/{\sqrt{-g}})({\delta S_{\rm matter}}/{\delta g_\mn})$ is
the matter stress-energy tensor, and 
\begin{equation}
\label{khSET}
\begin{split}
        &T^{{\rm kh}}_\mn\equiv \nabla_\r \left[{J_{(\m}}^\r u_{\nu)} -{J^\r}_{(\mu}u_{\nu)} -J_{(\mn)}u^\r\right]+\a\; a_\mu \, a_\nu\\
        &+\left(u_\s\, \nabla_\r J^{\r\s}-\a \,a_\rho a^\rho \right) u_\mu \, u_\nu +\frac{1}{2}L_\kh\; g_\mn+2 {\AE}_{(\mu}u_{\nu )}\,,
\end{split}     
\end{equation}
with
\begin{align}   
&{J^\r}_\mu\equiv \l\; (\nabla_\s u^\s)\; \d_\m^\r+\b\; \nabla_\m u^\r+\a\; a_\m u^\r\,,\\
&\AE_\m \equiv  (g_\mn - u_\mu\, u_\nu) \(\nabla_\r J^{\r\n}-\a \, a_\r \nabla^\n u^\r \)\,,\\
& L_\kh= \l\;(\nabla_\m u^\m)^2+\b\nabla_\m u^\n \nabla_\n u^\m+\a\; a_\m a^\m\,,
\end{align}
is the khronon stress-energy tensor. 

By varying with respect to $T$ while keeping $g^{\mu\nu}$ fixed, one obtains the khronon equation~\cite{Barausse:2012qh,Barausse:2013nwa}
\begin{equation}\label{khAetherEq}
        \nabla_\mu \left(\frac{\AE^\mu}{\sqrt{\nabla^\a T \nabla_\a T}}\right)=0\,.
\end{equation}
Note however that this equation follows from the modified Einstein equations~\eqref{khEinsteinEq}, the Bianchi
identity, and the equations of motion for matter
(which imply $\nabla_\m T^\mn_{\rm matter}=0$)~\cite{Jacobson:2010mx}. In the following, without any loss of generality, 
we will therefore solve
the modified Einstein equations \eqref{khEinsteinEq} alone.

The coupling constants $\a,\,\b$ and $\l$ must satisfy a number of theoretical requirements (absence of gradient instabilities and ghosts)~\cite{Blas:2011zd,Jacobson:2004ts,Garfinkle:2011iw}
and experimental constraints (absence of vacuum Cherenkov radiation~\cite{Elliott:2005va}, solar system experiments~\cite{Will:2014kxa,Blas:2011zd,Bonetti:2015oda}, gravitational wave propagation bounds from GW170817~\cite{Monitor:2017mdv,Gumrukcuoglu:2017ijh,ramos}, and cosmological constraints~\cite{Carroll:2004ai,Afshordi:2009tt,Yagi:2013ava,Yagi:2013qpa}). We refer the reader to Sec. II.A of Ref.~\cite{ramos} for more details, but 
for our purposes it suffices to recall that $\b$ is constrained to be very small ($|\b|\lesssim 10^{-15}$) by GW170817. Moreover,
as already mentioned in Sec.~\ref{intro}, by combining the aforementioned theoretical and experimental bounds (and particularly those from solar system tests and GW170817), one finds that either both $\a$ and $\l$ need to be small ($|\alpha|\sim |\lambda|\lesssim 10^{-5}$), or
$\a$ needs to be very small ($|\a|\simlt 10^{-7}$) while $\l$ is relatively unconstrained aside from
stability requirements~\cite{Blas:2009qj,Gumrukcuoglu:2017ijh} and  cosmological bounds~\cite{Carroll:2004ai,Afshordi:2009tt,Yagi:2013ava,Yagi:2013qpa} ($0 \leq\l\lesssim 0.01-0.1$).

Taking these bounds into account, it is tempting to set the ``small'' couplings $\a$ and $\b$ exactly to zero, and explore the phenomenological
consequences of a finite $\l$. An additional motivation for this choice is that Ref.~\cite{ramos} showed that black holes
moving slowly relative to the khronon present curvature singularities at the universal horizon unless $\a=\b=0$. 
In the following we will thus set $\a=\b=0$ and explore the effect of a finite $\l$ on the dynamics of stellar binary systems
(and particularly neutron star binaries). Note that the dynamics of neutron star binaries in khronometric theory,
as well as the gravitational wave generation from these systems, was
studied in Refs.~\cite{Yagi:2013ava,Yagi:2013qpa}, but in a region of parameter space that did {\it not} include the region $|\a|\simlt 10^{-7}$, $|\b|\lesssim 10^{-15}$
and $0\leq \l\lesssim 0.01-0.1$ currently favored after GW170817.

\section{Gravitational wave generation and violations of the strong equivalence principle}\label{GW}

The main bottleneck for predicting the motion and gravitational wave fluxes of binary systems
consists in computing the sensitivities of the binary components~\cite{Foster:2007gr,Yagi:2013ava,Yagi:2013qpa,1975ApJ...196L..59E,0264-9381-9-9-015,Will:1989sk,Barausse:2015wia,Yagi:2015oca,ramos}. 
Physically, the sensitivities parametrize violations of the strong equivalence principle, i.e.~violations of
the universality of free fall for self-gravitating bodies. In more detail, while 
test bodies are bound to follow spacetime geodesics because of the
action \eqref{khaction}, which presents no direct coupling (at tree level) between the matter fields and the \ae ther/khronon,
deviations from geodesic motion may appear beyond tree level, and notably for strongly self-gravitating objects (where the
large metric perturbations can mediate effective interactions between the \ae ther/khronon and matter).

This effect, well known in scalar-tensor theories (where it is often referred to as the ``Nordtvedt effect''~\cite{1975ApJ...196L..59E,PhysRev.169.1014,Nordtvedt:1968qs}) can be captured by adopting an effective point particle model
for the binary components, where the masses are {\it not} constant, but depend on the velocity relative to the \ae ther~\cite{Foster:2007gr,Yagi:2013ava,Yagi:2013qpa,ramos}. 
Physically, this parametrization accounts for the fact that the gravitational binding energy of a body (which in turn contributes to the total mass of the object) 
will in general depend on the \ae ther, 
since the latter appears in the gravitational action, c.f.~Eq.~\eqref{khaction}.
The sensitivities are then defined as~\cite{Foster:2007gr,Yagi:2013ava,Yagi:2013qpa,ramos}
\be
\sigma\equiv -\frac{\dd\ln \tilde{m}(\gamma)}{\dd \ln\gamma}\Big|_{\gamma=1}\,,
\ee
where $\tilde{m}(\gamma)$ is the object's mass, depending on 
its Lorentz factor in the preferred frame (i.e. $\gamma\equiv {\bf U}\cdot{\bf u}$, where
${\bf U}$ is the body's four velocity). The sensitivities are then expected to be small
for objects with weak internal gravity (for which the gravitational binding energy, and thus the \ae ther contribution to the mass, are negligible), and must vanish in the test particle limit (i.e.~in the limit in which the self-gravity of the body is neglected).

In general, non-vanishing sensitivities impact the dynamics of a binary system in both the conservative and dissipative sectors. At Newtonian order,
the acceleration of a binary component becomes~\cite{Foster:2007gr,Yagi:2013ava,Will:2018ont}
\begin{equation}
\label{eq:New_active}
\dot{v}_A^i= -\frac{{\cal G} {m}_B \hat n_{AB}^i}{r_{AB}^2}\,,
\end{equation}
where $A,B$ range from 1 to 2 and denote the two bodies,  $r_{AB}=|\boldsymbol{x}_A-\boldsymbol{x}_B|$, $\hat n_{AB}^i=(x^i_A-x^i_B)/r_{AB}$, and  we have defined the {\it{active}} gravitational masses
\be
\label{active-mass}
m_B\equiv (1+\sigma_B) \tilde{m}_B({\gamma_B=1})
\ee
and the ``strong field'' Newton constant
\be
\label{calG}
{\cal G}\equiv\frac{G_N}{(1+\sigma_A) (1+\sigma_B)}\,.
\ee
The sensitivities also appear in the binary acceleration at 1PN order~\cite{Foster:2007gr,Will:2018ont}, together with their derivatives with respect to $\gamma$,
which one can parametrize with the additional parameters~\cite{Foster:2007gr,Yagi:2013ava}
\begin{equation}
 \sigma' \equiv \sigma + \sigma^{2} + \left.\frac{d^{2} \ln \tilde{m}(\gamma)}{d (\ln \gamma)^{2}}\right|_{\gamma = 1}\,.
\end{equation}

Similarly, the sensitivities appear in the dissipative sector, e.g the energy flux emitted by a quasi-circular binary in gravitational waves is given by~\cite{Foster:2007gr,Yagi:2013ava}
\vskip 0.2cm
\begin{align}
\label{Pdot-AE}
\frac{\dot{E}_{b}}{E_{b}} =  \displaystyle 2  & \left(\frac{ {\cal G} G m_1 m_2}{r_{12}^3}\right)
  \Bigg\{ \frac{32}{5}({\cal A}_1+{\cal S}{\cal A}_2+ {\cal S}^2{\cal A}_3) v_{12}^2 \nonumber \\ 
  &\displaystyle+\left(s_1-s_2\right)^2\Bigg[{\cal C}+\frac{18}{5}{\cal A}_3\, V^2_{CM}\nonumber\\&+\left(\frac{6}{5}{\cal A}_3+36 {\cal B}\right) (\boldsymbol{V}_{CM}\cdot\boldsymbol{n}_{12})^2\Bigg]  
   \nonumber
 \Bigg\}\,,
\end{align}
where
\be
E_{b} = - \frac{{\cal{G}} m_1 m_2}{2 r_{12}}\,,
\ee
 is the binary's total energy (potential and kinetic); $V_{CM}$ is the velocity of the center of mass relative to the preferred frame; and we
have defined  the rescaled sensitivities
\be
s_A\equiv\frac{\sigma_A}{1+\sigma_A}\,,
\ee
as well as the coefficients
\begin{align}
\label{A1-def-HL}
{\cal A}_1 &\equiv \frac{1}{c_2}+\frac{3 \a({\cal Z}-1)^2}{2 c_0(2-\a)}, \quad {\cal A}_2\equiv \frac{2({\cal Z}-1)}{(\a-2)c_0^3}, \\ 
\label{A3-def-HL}
{\cal A}_3 &\equiv \frac{2}{3\a(2-\a)c_0^5}, \quad {\cal B} \equiv \frac{1}{9\a \,c_0^5(2-\a)},\\ 
{\cal C} &\equiv \frac{4}{3 c_0^3\,\a(2-\a)},\quad {\cal S}\equiv s_1\, \frac{m_2}{m}+s_2\,\frac{m_1}{m}, \\ 
{\cal Z}&\equiv \frac{(\a_1-2\a_2)(1-\b)}{3(2\b-\a)}\,,
\end{align}
which depend in turn on the total mass $m=m_1+m_2$, on the spin-0 and spin-2 propagation speeds
\begin{subequations}\label{kh speeds}
\begin{align}
c_2^2=&\frac{1}{1-\b}\,,\\
c_0^2=&\frac{(\lambda+\beta)(2-\alpha)}{\alpha (1-\beta) (2+3\lambda+\beta)}\label{spin0speed}
\end{align}
\end{subequations}
and on the preferred-frame parameters~\cite{Blas:2011zd,Bonetti:2015oda}
\begin{align}
\label{alpha1-HL}
\alpha_1 &=  \frac{4(\alpha-2 \beta)}{\beta-1}\,,
\\ 
\label{alpha2-HL}
\alpha_2 &=
\frac{(\alpha-2\beta)[-\beta^2+\beta(\alpha-3)+\alpha+\lambda(-1-3\beta+2\alpha)]}{(\beta-1)(\lambda+\beta)(\alpha-2)}\,.
\end{align}

Note that dipole gravitational wave emission is proportional to  ${\cal C}$ and to $(s_1-s_2)^2$ (as in scalar-tensor theories~\cite{1975ApJ...196L..59E,0264-9381-9-9-015,Will:1989sk}), and that it
may dominate over quadupole emission at the low frequencies relevant for binary pulsars, depending on the sensitivities and 
the theory's coupling parameters.
However, in the limit $\a,\b\to 0$ with $\l\neq0$, the spin-0 propagation speed diverges (i.e. the spin-0 mode becomes non-propagating)
and one has ${\cal A}_1\to1$ and ${\cal A}_2,\,{\cal A}_3,\,{\cal B},\, {\cal C}\to0$. As a result, there exists no 
dipole gravitational emission in this limit, while quadrupole fluxes match exactly those of general relativity. This means that the damping of the period
of binary pulsar systems is not suitable for constraining khronometric theory in this limit. 

Nevertheless, 
timing observations of pulsar systems also constrain modifications
to the conservative sector.
For instance, in binary systems
the acceleration depends on the sensitivities $\sigma_A$ at Newtonian order, and on $\sigma_A$ and their ``derivatives''
$\sigma_A'$ at 1PN order~\cite{Foster:2007gr,Will:2018ont}. As a result, observations of the 1PN dynamics (e.g.~periastron precession and preferred frame effects) of binary pulsar systems
can in principle constrain khronometric theory~\cite{Foster:2007gr,Will:2018ont}, provided that one can compute  $\sigma_A$ and  $\sigma_A'$ and relate them to the theory's coupling constants.
Note that binary pulsar
observations do not constrain the dependence of the Newtonian acceleration on the sensitivities [Eq.~\eqref{eq:New_active}], since the Newtonian dynamics of the system is used to infer the masses of the binary components [i.e. the 
sensitivities are re-absorbed into the mass redefinition given by Eq.~\eqref{active-mass}], and one has therefore to
resort to the 1PN dynamics to test the theory. However, direct bounds on the Newtonian acceleration (and thus on the sensitivities) 
are provided by the 
triple system PSR~J0337+1715~\cite{Archibald:2018oxs}, which consists of a pulsar-white dwarf binary system, and a second white dwarf farther out.
Timing of this system allows one to constrain the strong equivalence principle, i.e. to 
assess whether the two binary components fall with different accelerations toward the third body. 
The fractional acceleration difference $\Delta$ between the binary components is constrained 
to $|\Delta|<2.6\times 10^{-6}$~\cite{Archibald:2018oxs}. This can in turn be translated into
a bound on the sensitivities by using Eq.~\eqref{eq:New_active} (c.f. also Ref. \cite{Will:2018ont}).
In more detail, since the sensitivities are expected to scale with the gravitational binding energy of the star~\cite{Foster:2007gr,Yagi:2013ava},
and thus to be much smaller for white dwarfs than for neutron stars, from Eq.~\eqref{eq:New_active} one obtains $|\Delta|=|\sigma_{\rm p}|+{\cal O}(\sigma_{\rm p})^2$, where $\sigma_{\rm p}$ is the pulsar's sensitivity.

\section{The ans{\"a}tze for the metric and khronon of slowly moving stars}
\label{sec_ans}
As shown in Refs.~\cite{Foster:2007gr,Yagi:2013ava,ramos}, the sensitivities can be extracted from solutions to the field equations 
describing an object moving slowly relative to the preferred foliation (as defined far from the body). In more detail, let us consider a compact object (e.g. a neutron star or a black hole)
at rest in the stationary flow of a khronon field
moving with speed $-v$ along the $z$-direction on flat space near spatial infinity.
The most generic metric and \ae ther ans{\"a}tze for such a system are given, through order $\mathcal{O}(v)$ and in Eddington-Finkelstein coordinates, by~\cite{ramos}
\begin{align}
\label{metric_Ansatz0}
{g}_{\mu\nu} dx^\mu dx^\nu =
& f(r) \dv^2 -2B(r) \dr \dv-r^2 \dOm^2\nn\\
&+v\,\Big\{{\dv}^2 f(r)^2 \cos \theta \psi (r)\nn\\
&-2 {\dd\theta } {\dr} \sin \theta [\Sigma (r)-B(r) \chi (r)]\nn\\
&+2 {\dr}  {\dv} f(r) \cos \theta [\delta (r)-B(r) \psi (r)]\nn\\
&+{\dr}^2 B(r) \cos \theta [B(r) \psi (r)-2 \delta (r)+2 \Delta (r)]\nn\\
&-2 {\dd\theta } {\dv} f(r) \sin \theta \chi (r)\Big\} +\mO(v^2)\,, 
\end{align}
and
\begin{align}\label{aether_Ansatz0}
{u}_\mu \dx^\mu=&\bar{u}_\vv(r)\dv -A(r)B(r)\dr+v\,\Bigg\{\frac{1}{2} f(r) \cos \theta \times\nn\\
& \Bigg[2 \bar{u}^r(r) \left(\frac{B(r) \Delta (r) \bar{u}^r(r)}{\bar{u}_\vv(r)}+\delta (r)-\eta (r)\right) \nn\\
   &+\psi (r) \bar{u}_\vv(r)\Bigg] \dv+\frac{1}{2} \cos \theta \times \nn\\
   &\Bigg[B(r) \Big(-\frac{2 B(r) \Delta (r) \bar{u}^r(r)^2}{\bar{u}_\vv(r)}-2 \delta (r) \bar{u}^r(r)\nn\\
   &-\psi (r) \bar{u}_\vv(r)\Big)+2 A(r)f(r) \eta (r)\Bigg] \dr\nn\\
   &-\sin \theta \Pi (r) \bar{u}_\vv(r) d\theta\Bigg\}+\mO(v^2)\,,
\end{align}
where $\bar{u}_\vv=(1+fA^2)/(2A)$ and $\bar{u}^r=(-1 + A^2 f)/(2 AB)$ are the background [i.e. $\mO(v^0)$] \ae ther components.
Hypersurface orthogonality of the \ae ther then requires~\cite{ramos} 
\be\label{eta}
\begin{split}
\eta(r)=&-\frac{2\bar{u}^r(r)^3 B(r)^3 \Delta (r) -2 \bar{u}_\vv(r)^3 \Pi '(r)}{2 f(r) \bar{u}_\vv(r)}\\
&-\frac{B(r)^2 \bar{u}_\vv(r) \bar{u}^r(r) \left[2 \delta(r) \bar{u}^r(r)+\psi (r) \bar{u}_\vv(r)\right]}{2 f(r) \bar{u}_\vv(r)}\,.
\end{split}
\ee

Considering  an infinitesimal gauge transformation with generator 
$\xi^\mu \partial_\mu=v\Omega(r)(-r  \cos\theta\partial_r+ \sin\theta\partial_\theta)+v K(r) \cos\theta \partial_{\rm v}$, one can set $\Pi=\Delta=0$ by tweaking $\Omega$ and $K$ and re-defining
the other free functions appearing in the ansatz~\cite{ramos}.
Note also that by requiring the spacetime to be asymptotically flat [i.e. $g_{\mu\nu}=\eta_{\mu\nu}+{\cal O}(1/r)$]
and the \ae ther to asymptote to $u^\mu\partial_\mu= \partial_t-v \partial_z+{\cal O}(v)^2$ in suitably defined ``Cartesian`'' coordinates $(t,x,y,z=r \cos\theta)$
near spatial infinity, one finds that 
 the potentials in the ansatz above must satisfy the boundary conditions
$\psi,\Sigma \to 0$, $\delta\to-1$ and $\chi/r\to -1$ for $r\to \infty$~\cite{ramos}.\footnote{This is
easy to see in the following manner. These conditions, inserted into the ans{\"a}tze \eqref{metric_Ansatz0} and
\eqref{aether_Ansatz0}  lead to $\ds^2\approx \dt^2-\dr^2-r^2\dd\Omega^2-2v \dt \dz$ and $u_\mu dx^\mu\approx dt$,
where we have changed the time coordinate to $t\approx\vv-r$ and defined $z=r\cos\theta$.
Further changing the time coordinate to $t'=t-vz$ yields the flat line element, while the \ae ther transforms to
${u}^\m \partial_\mu\approx \partial_{t'}-v\partial_z$ asymptotically, i.e. at spatial infinity the \ae ther moves with velocity $-v$
with respect to the flat asymptotic metric.}

In more detail, the perturbative solution to the field equations \eqref{khEinsteinEq} near spatial infinity that satisfies these boundary conditions is~\cite{ramos}
\begin{subequations}\label{asymptotic}
\begin{align}
\d(r)=&\,-1-\frac{2(\b+\l)(-r_s+2\c_0)}{(1-3\b-2\l)r}+\mO\left(\frac{1}{r^2}\right)\,,\label{asymptotic delta}\\
\c(r)=&\,- r+\c_0+\mO\left(\frac{1}{r}\right)\,, \label{asymptotic chi}\\
\psi(r)=&\,\frac{-3\b (3-2a_2)r_s^2-2\S_1}{3\, r^2}+\mO\left(\frac{1}{r^3}\right)\,,\\\
\S(r)=&\,\frac{\S_1}{r}+\mO\left(\frac{1}{r^2}\right) \,,\\
f(r) =& 1-\frac{2r_s}{r}- \frac{\a r_s^3}{6r^3}+ \mO\left(\frac{1}{r^4}\right) \label{asyF}\\
B(r) =& 1+\frac{\a r_s^2}{4r^2}+\frac{2\a r_s^3}{3r^3}+\mO\left(\frac{1}{r^4}\right) \label{asyB} \\
A(r) =& 1+ \frac{r_s}{r}+\frac{a_2  r_s^2}{r^2}+\left(24 a_2+\a-6\right) \frac{r_s^3}{12 r^3}+\mO\left(\frac{1}{r^4}\right) \label{asyA}\,,
\end{align}
\end{subequations}
where $r_s=G_N M$ (with $M$ the object's mass), while
$\c_0$, $\c_2$, $\Sigma_1$ and $a_2$ are parameters that describe the object,
and which in practice are to be extracted from the full (and typically numerical) ``strong-field'' solution~\cite{ramos,Yagi:2013ava}. 
The sensitivities can then be read off this solution, and in particular
from $\c_0$, as~\cite{ramos}
\be\label{sensitivity}
\begin{split}
\sigma= &\frac{\a-\b-3\a \b+5\b^2+\l-2\a \l +3\b \l }{(2-\a)(1-3\b-2\l)} \\&-\frac{2(1-\b)(\b+\l)}{(2-\a)(1-3\b-2\l)}\frac{\c_0}{r_s}\,.
\end{split}
\ee

For a star, unlike for a black hole, the background is static, i.e.~one has $A=1/\sqrt{f}$ and $\bar{u}^r=0$~\cite{Eling:2007xh,Yagi:2013ava},
and it is convenient to express the metric and \ae ther in Schwarzschild coordinates, which
are related to the Eddington-Finkelstein coordinates used above and in Ref.~\cite{ramos} by the simple transformation
$\dv = {\rm d}t + B(r)/f(r) {\rm d}r$. 
Let us also use spatial coordinates comoving with the fluid, i.e.
let us propagate the spatial coordinates of the initial spatial hypersurface along the fluid's worldlines, as discussed in Ref.~\cite{Yagi:2013ava},
so that the fluid 4-velocity is simply $\boldsymbol{U}_{\rm fluid}\propto \partial_{\rm v}=\partial_t$.
Note that this coordinate choice
is compatible with the gauge transformations used above to set  $\Pi=\Delta=0$ (which consist indeed of a
time independent transformation of the spatial coordinates -- thus equivalent to relabeling the spatial coordinates on the 
initial foliation -- and an infinitesimal redefinition of the time coordinate $\rm v$). This can also be checked explicitly by
noting that $\boldsymbol{U}_{\rm fluid}\propto \partial_{\rm v}=\partial_t$ is unaffected by a change of coordinates with generator $\xi^\mu \partial_\mu=v\Omega(r)(-r  \cos\theta\partial_r+ \sin\theta\partial_\theta)+v K(r) \cos\theta \partial_{\rm v}$, which one needs to set $\Pi=\Delta=0$.

At this point, unlike for a black hole, we can require that the ans{\"a}tze for the metric and \ae ther 
be left unchanged by the joint transformations 
$t\to-t$ and $v\to-v$, i.e. by a joint reflection of the (Schwarzschild) time and of the velocity~\cite{Yagi:2013ava}.
It can easily be checked that this requires $\Sigma=\psi=0$, which in turn implies $\eta=0$ due to the orthogonality condition \eqref{eta}. Note that this requirement
cannot be imposed for black holes since the background \ae ther has a radial component, i.e. a time reflection
transforms the \ae ther from ingoing to outgoing~\cite{Barausse:2011pu,ramos}. Similarly, invariance under a 
joint reflection of time and velocity implies that perturbations to the density and pressure of the fluid should vanish at 
$\mathcal{O}(v)$~\cite{Yagi:2013ava}.

In summary, the metric, \ae ther and fluid ans{\"a}tze for a star therefore become, in Schwarzschild coordinates,
\begin{align}
\label{metric_Ansatz}
&{g}_{\mu\nu} dx^\mu dx^\nu =
 f(r) {\rm d}t^2 -\frac{B(r)^2}{f(r)} \dr^2 -r^2 \dOm^2\nn\\
&+2v f(r) {\rm d}t\,[  \cos \theta \delta (r) {\dr} 
- \sin \theta \chi (r) {\dd\theta } ] +\mO(v^2)\,, 
\end{align}
\begin{equation}\label{aether_Ansatz}
{u}_\mu \dx^\mu={\sqrt{f(r)}} {\rm d} t+\mO(v^2)\,,\quad U_{\rm fluid}^\mu  \partial_\mu=\frac{\partial_t}{\sqrt{f(r)}}+\mO(v^2)\,,
\end{equation}
\be
\rho(r)=\bar{\rho}(r)+\mO(v^2)\,,\quad p(r)=\bar{p}(r)+\mO(v^2)\,,
\ee
where overbars denote background $\mO(v^0)$ quantities. 

Note that these ans{\"a}tze, while obtained in a slightly different way, reduce to those successfully used by Ref.~\cite{Yagi:2013ava} to compute neutron star 
sensitivities for generic $\a$, $\b$ and $\l$, with the additional ``dipole'' assumption
$V(r,\theta)\propto \delta(r) \cos\theta$ and $S(r,\theta)\propto \chi(r)\sin\theta$. This dipolar structure corresponds to taking $n=1$ in Eqs. (145), (146) and (147) of
Ref.~\cite{Yagi:2013ava}, and indeed it is from those $n=1$ equations that Ref.~\cite{Yagi:2013ava} extracts neutron star sensitivities.

\section{Solutions for slowly moving stars and their sensitivities}\label{ext}
For $\alpha=\beta=0$, the stellar exterior background solution is given by $f(r)=1-2 r_s/r$ and $B(r)=1$~\cite{Berglund:2012bu,ramos}. 
The field equations for the potentials $\delta$ and $\chi$ reduce to the homogeneous coupled system
\begin{align}
\chi''(r)=&\frac{1}{{r (2 r_s-r)}} \{\delta '(r)(1+2\l) r (2 r_s-r)\nn\\&+2 \delta (r) [(\lambda -2) r_s-2
   \lambda  r]+4 r_s \chi '(r)+4 \lambda 
   \chi (r)\}\,,\label{chi2}\\
\delta''(r)=&\frac{1}{\lambda  r^2 (r-2 r_s)^2}\{\delta (r) [5 \lambda  r_s^2+2(1-4 \lambda ) r_s r\nn\\&
+(2 \lambda -1) r^2]+r
   (2 r_s-r) [2 \lambda  (r_s+r) \delta '(r)\nn\\&-(2 \lambda +1) \chi '(r)]+2
   \lambda  (5 r_s-2 r) \chi (r)\}\,.
\end{align} 
Let us define, without loss of generality, $\delta(r)=\chi'(r)+\lambda h(r)$. This redefinition makes the equations above of
third order in the derivatives of $\chi$, but the $\chi'''(r)$ terms can be eliminated by using the derivative of Eq.~\eqref{chi2}, thus yielding
\begin{align}
&\chi''(r)=\frac{1}{2 r (2 r_s-r)}\{(2 \lambda +1) r (r-2 r_s) h'(r)\nn\\&+h(r) [4 \lambda  r-2 (\lambda -2) r_s]-2 (r_s-2 r)
   \chi '(r)-4 \chi (r)\}\label{chi2bis}\,,\\
&h''(r)=\frac{1}{r^2 (r-2 r_s)^2}[7 r_s r (2 r_s-r) h'(r)\nn\\&\qquad\qquad+2 h(r) \left(-10 r_s^2+2 r_s r+r^2\right)]\,.\label{h2}
\end{align}
The second equation can be solved analytically to give
\be\label{hsol}
h(r)=\frac{k_1 r^2 \left(3 r_s^2-6 r_s r+2 r^2\right)}{3 r_s^3 (r-2 r_s)^2}+\frac{k_2
   r^{5/2}}{\sqrt{r-2 r_s}}\,,
\ee
with $k_1$ and $k_2$ integration constants. 

From the boundary conditions at spatial infinity, $\delta\to-1$ and $\chi/r\to -1$ for $r\to \infty$, it follows that it must be
$k_2=-2 k_1/(3 r_s^3)$. For this choice of the integration constants, the equation for $\chi$ [Eq.~\eqref{chi2bis}]
can be solved analytically with Lagrange's method of the variation of constants. The explicit solution
is quite cumbersome and hardly enlightening, so we will only present here its perturbative expansion near spatial infinity,
which reads
\begin{multline}\label{chi_pert}
\chi(r)=\frac{C_0 r}{r_s} + \frac{k_1 - 2 \lambda k_1 - 6 C_0}{12}
+\frac{(7 - 4 \lambda) k_1 r_s}{24 r}\\+ \frac{(2 + 3 \lambda) k_1 r_s^2 \ln(r/r_s)}{9 r^2}
+\frac{C_1 r_s^2}{r^2}
+{\cal O}\left(\frac{\ln r}{r^3}\right) \,,
\end{multline}
with $C_0$ and $C_1$  integration constants. Since the system given by Eqs.~\eqref{chi2bis} and \eqref{h2}
is homogeneous, its solutions are invariant under  a global rescaling, i.e. if the pair $(h,\chi)$ is a 
solution,  $(\Lambda h,\Lambda\chi)$, with $\Lambda$ an arbitrary constant, is also a solution.
Rescaling therefore Eq.~\eqref{chi_pert} by a factor $-r_s/C_0$ [in order to match Eq.~\eqref{asymptotic chi}],
one can then compute the sensitivity, once the exterior solution for $\chi$ and $h$ is known, via
Eq.~\eqref{sensitivity}. Setting $\a=\b=0$ and [from Eq.~\eqref{chi_pert}] $\chi_0=-r_s (k_1 - 2 \lambda k_1 - 6 C_0)/(12 C_0)$, one obtains
\begin{equation}\label{eq_sens}
\sigma=\frac{\lambda k_1}{12 C_0}\,.
\end{equation}

As a sanity check, note  that for the special choice $k_1=0$, one gets $\sigma=0$. This makes perfect sense, 
because $k_1=0$ implies $h(r)=0$ and thus
$\delta(r)=\chi'(r)$. An infinitesimal gauge transformation $\vv'=\vv+v\;\chi(r)\,\cos\theta+{\cal O}(v)^2$ can then be used to
eliminate all ${\cal O}(v)$ terms in the metric \eqref{metric_Ansatz}, i.e. that metric
becomes simply (in the exterior) the Schwarzschild metric in Eddington-Finkelstein coordinates [since for $\alpha=\beta=0$ one has
$f(r)=1-2r_s/r$ and $B(r)=1$].
In general, however, $k_1$ may {\it not} be zero. In fact, 
the exact values of the integration constants $k_1$ and $C_0$ must be obtained by matching the (analytically known) exterior solution
for $h$ and $\chi$ to the stellar interior, at the surface of the star. This was indeed the procedure followed in Refs.~\cite{Yagi:2013qpa,Yagi:2013ava}.

The field equations for $h$ and $\chi$ in the presence of a perfect fluid source are only slightly more involved than
the vacuum system \eqref{chi2bis}--\eqref{h2}. To derive them, one can first note that the field equations of khronometric theory
for static spherically symmetric stars only differ from their general-relativistic counterparts if $\a\neq0$~\cite{Eling:2007xh,Yagi:2013qpa,Yagi:2013ava}.
Since we are considering here the case $\a=\b=0$, the background quantities $f(r)$, $B(r)$, $\bar{\rho}(r)$ and $\bar{p}(r)$
will therefore be the same as in general relativity, i.e. they can be easily computed by solving the Tolman–Oppenheimer–Volkoff equations~\cite{tov1,tov2} while imposing regularity at $r=0$.
The field equations for $h$ and $\chi$ can then be obtained by the same procedure as in vacuum, 
i.e.~replace the ansatz $\delta(r)=\chi'(r)+\lambda h(r)$ in the ${\cal O}(v)$
equations, and use the derivatives of the equations themselves to eliminate the $\chi'''$ terms. Making also repeated use of the 
${\cal O}(v)^0$ equations (i.e.~the Tolman–Oppenheimer–Volkoff equations) and their derivatives, after some algebra one obtains
\begin{align}\chi''(r)&=-\frac{1}{2 r^2 f(r)}\Big\{B(r)^2 \Big[r h(r) \Big(3 \lambda +4 \pi  (4 \lambda +3) r^2 G\bar{p}(r)\nn\\&-4 \pi  (2
   \lambda +1) r^2 G\bar{\rho} (r)+2\Big)\nn\\&+r \Big(16 \pi  r^2 G\bar{p}(r)-8 \pi  r^2 G\bar{\rho}
   (r)+3\Big) \chi '(r)-4 \chi (r)\Big]\nn\\&+r f(r) \Big[(2 \lambda  +1) r
   h'(r)+(\lambda -2) h(r)+\chi '(r)\Big]\Big\}\,,\label{chi2int}\\
h''(r)&=\frac{1}{2 r^2 f(r)^2}\Big\{B(r)^2 f(r) \Big[r h'(r) \Big(-32 \pi  r^2 G\bar{p}(r)\nn\\&+24 \pi  r^2 G\bar{\rho}
   (r)-7\Big)\nn\\&+8 h(r) \Big(\!\!-\!\pi  r^2 G\bar{p}(r)+\pi  r^3 G\bar{\rho} '(r)-5 \pi  r^2 G\bar{\rho}
   (r)+2\Big)\Big]\nn\\&+2 B(r)^4 h(r) \Big[8 \pi  r^2 G\bar{p}(r)+1\Big] \Big[8 \pi  r^2
   G\bar{\rho} (r)-1\Big]\nn\\&+f(r)^2\Big[7 r h'(r)-10 h(r)\Big]\Big\}\label{h2int}
\end{align}
Remarkably, just like in vacuum, the equation for $h$ does  not depend on $\chi$.

Like the Tolman–Oppenheimer–Volkoff equations, Eqs.~\eqref{chi2int}--\eqref{h2int}  have a singular point at $r=0$. In principle, we should thus solve
these equations perturbatively near the center of the coordinates, imposing regularity by requiring that
$h$ and $\chi$ (as well as $\bar{p}$, $\bar{\rho}$, $F$ and $B$) are analytic functions near $r=0$. 
As we will show below, the general regular perturbative solution for $h$ and $\chi$ depends on two integration constants,
one of which can be fixed by exploiting the homogeneity of the system~\eqref{chi2int}--\eqref{h2int}, i.e.
the fact that if $(h,\chi)$ is a 
solution for the interior geometry,  $(\Lambda h,\Lambda\chi)$, with $\Lambda$ an arbitrary constant, is also a solution.
One may then use the perturbative solution, which now depends on only
one undetermined integration constant, to ``move away'' from the center, i.e. to provide initial conditions 
at small $r\neq0$ for an outward numerical integration. 
One could then try to tune that undetermined integration constant to match the numerical interior solution to the  
asymptotically flat exterior at the stellar surface, e.g. by a shooting method (as was done 
in similar problems in Refs.~\cite{Barausse:2011pu,Barausse:2015frm,ramos,Yagi:2013ava,Yagi:2013qpa}).
However, while counting the degrees of freedom of the perturbative solution
would seem to allow for this procedure, the latter is doomed to fail in our particular case, as we will now explain, because of the structure of the system~\eqref{chi2int}--\eqref{h2int}.

Indeed, let us look in more detail at the perturbative solution for $h$ near the origin.
To obtain it, one needs first to solve the background Tolman–Oppenheimer–Volkoff equations imposing regularity at $r=0$, i.e.
with the ans{\"a}tze $\bar{p}(r)=\sum_{n=0}^{+\infty} \bar{p}_{2n} r^{2n}$,  $f(r)=\sum_{n=0}^{+\infty} f_{2n} r^{2n}$
and  $B(r)=\sum_{n=0}^{+\infty} B_{2n} r^{2n}$ (with the odd powers set to zero by the field equations,
and with the density expressed in terms of the pressure via a suitable equation of state). The background
solution can then be inserted into Eq.~\eqref{h2int}, and one can look 
for a solution of the form $h(r)=\sum_{n=0}^{+\infty} h_{2n} r^{2n}$ (note that the field equations forbid again the presence of odd powers).
This procedure yields 
\begin{equation}\label{sol_h_pert}
h(r)=h_2 \left[r^2+\frac{2}{15} \pi  r^4  (7 G\bar{\rho}_0-33 G\bar{p}_0)+ {\cal O}(r^6)\right]\,,
\end{equation}
where $\bar{\rho}_0$ and $\bar{p}_0$ are the central density and pressure, $h_2$ is an undetermined integration constant, while
the higher order terms are
proportional to $h_2$ and are otherwise completely determined by the perturbative background solution
for $f$, $B$, $\bar{p}$ and $\bar{\rho}$ near the center (for which we have chosen a gauge in which $f(0)=1$). 
Therefore, the perturbative
regular solution for $h$ near the origin has only one integration constant, $h_2$. 

Because of the  homogeneity of the system~\eqref{chi2int}--\eqref{h2int}, 
we can then rescale $h_2=1$, in which case the solution for $h$ near the center is completely determined.\footnote{Note that rescaling $h_2=1$
is justified as it corresponds to dividing the original system~\eqref{chi2int}--\eqref{h2int} 
by $h''(0)/2$. One can then introduce new variables $h_{\rm new}=h/(h''(0)/2)$ and
$\delta_{\rm new}=\delta/(h''(0)/2)$. Because of the linearity and homogeneity of the system, the resulting equations will then still be given by Eqs.~\eqref{chi2int}--\eqref{h2int}, but with $h\to h_{\rm new}$ and $\delta\to \delta_{\rm new}$ and with the extra constraint $h_{\rm new}''(0)=2$. This latter condition then forces $h_2$ in Eq.~\eqref{sol_h_pert}.
} Therefore, even though 
 solving Eq.~\eqref{chi2int} for $\chi$ perturbatively near $r=0$ will yield a second integration constant for the system~\eqref{chi2int}--\eqref{h2int} (i.e.
the general solution turns out to be $\chi(r)=\sum_{n=0}^{+\infty} \chi_{2n+1} r^{2n+1}$ with all the coefficients $\chi_{2n+1}$ given
in terms of $h_2$ and $\chi'(0)=\chi_1$), 
there is not enough freedom (at least for generic matter equations of state)
 to match the interior solution for $h$ to an asymptotically flat exterior [i.e. to Eq.~\eqref{hsol} with $k_2=-2 k_1/(3 r_s^3)$].

Alternatively, another equivalent way of understanding why the matching is not possible is the following. One is of course free to keep $h_2$ generic
(rather than set it 1 as done above), but that will only provide a global 
rescaling of the interior solution [as is obvious from the homogeneity of the system \eqref{chi2int}--\eqref{h2int}]. Similarly, the exterior asymptotically flat solution
given by Eq.~\eqref{hsol} with $k_2=-2 k_1/(3 r_s^3)$ depends on just one integration constant ($k_1$), which is
also a global amplitude because of homogeneity. One may then try to match the two solutions by imposing continuity of
$h$ and $h'$ at the star's surface, which will yield two conditions $h_2 H_1= k_2 H_2$ and $h_2 H_3= k_2 H_4$, where $H_1$, $H_2$, $H_3$
and $H_4$ depend on the interior and exterior solutions and their derivatives evaluated at the matching point. Clearly, this system
can only be satisfied if $H_2/H_1=H_4/H_3$. Since the interior solution will depend on the matter equation of state (e.g. for a neutron star, the equation of state of  nuclear matter), 
such a ``miracle'' cannot happen for generic equations of state.

More in general, note that the matching turns out to be impossible because the equation for $h$ does  not depend on $\chi$, i.e. if the equation for $h$
were coupled to $\chi$, the perturbative solutions for $h$ and $\chi$ may both depend on the same two integration constants,
which would leave us with enough freedom to match to the exterior solution. We have verified this also by changing the
interior equation \eqref{h2int} ``by hand'', introducing an artificial source term depending on $\chi$ on the right-hand side.

In any case, as a result,
we obtain that for generic equations of state  the only solution for $h$ that is regular at the center and asymptotically flat is the trivial one, $h(r)=0$.
 This thus yields vanishing sensitivities by virtue of Eq.~\eqref{eq_sens}. Moreover,  since $h(r)=0$ implies that
 $\delta=\chi'$, one can perform an infinitesimal gauge transformation $\vv'=\vv+v\;\chi(r)\,\cos\theta+{\cal O}(v)^2$ to
eliminate all ${\cal O}(v)$ terms in the metric \eqref{metric_Ansatz}, i.e. not only do the sensitivities vanish, but the
entire geometry for slowly moving stars matches that of the corresponding general relativistic solution when $\a=\b=0$ and $\l\neq0$.

We stress that even though the solution for $h$ is trivial, that for $\delta=\chi'$ is not. Indeed, the solution for $\chi$ can be obtained explicitly
by setting $h(r)=0$ in Eq.~\eqref{chi2int}, and then by integrating numerically the resulting equation outward from the center of the star (imposing regularity there and using the homogeneity of the equation to set the one undetermined constant appearing in the perturbative solution -- $\chi_1$ -- to a given value, e.g. $\chi_1=1$). The integration can proceed through the stellar surface and then in the exterior [where the solution, as mentioned, can be  obtained also analytically
by Lagrange's method of the variation of constants, c.f. Eq.~\eqref{chi_pert} for its series expansion]. 
The fact that $\delta=\chi'$ is not trivial was of course to be expected from the boundary conditions ($\delta=\chi'\to -1$ at spatial infinity),
which correspond to an asymptotically flat geometry and {\it moving} \ae ther. In other words, for $\a=\b=0$ and $\l\neq0$ the geometry of
the star is the same as in general relativity, but the \ae ther is non-trivial (and in particular it corresponds to a flow with velocity $-v$ at spatial infinity).
This is the same situation that Ref.~\cite{ramos} found for black holes when  $\a=\b=0$ and $\l\neq0$, although
the non-trivial \ae ther configurations differ in the two cases. Like for black holes, however, it is clear that
the \ae ther stress energy tensor vanishes at order ${\cal O}(v)$, so that the metric matches the general relativistic solution.
In this sense the \ae ther behaves as a ``stealth'' field on the spacetime of stars moving slowly relative to the preferred foliation, when $\a=\b=0$.

\section{Discussion and conclusions}\label{disc}
We have studied the structure of slowly moving stars in khronometric theory, i.e. the low-energy limit of Ho\v rava gravity, in the region of parameter space
favored by experimental tests (and notably by solar system tests and the multimessenger bounds on gravitational wave propagation from GW170817)
and by theoretical considerations (and namely by the results of Ref.~\cite{ramos}, which found that two of the
dimensionless coupling constants of the theory -- $\a$ and $\b$ -- need to vanish exactly for black holes in motion relative to the preferred frame
to present regular universal horizons).

Solutions for stars moving slowly relative to the preferred frame are necessary to calculate the ``sensitivities'' -- i.e. the parameters
characterizing the effective coupling of the star to the Lorentz-violating scalar degrees of freedom of the theory -- and therefore assess possible violations
of the strong equivalence principle. Non-zero sensitivities may generally impact the dynamics of binary systems both in the conservative and dissipative sectors (e.g. they can trigger dipole gravitational emission, as well as deviations away from the quadrupole formula and the conservative Newtonian and PN dynamics of general relativity).

While neutron star sensitivities had previously been computed in Refs.~\cite{Yagi:2013ava,Yagi:2013qpa}, the parameter space considered in those works did not include the
case $\a=\b=0$ currently favored by GW170817 and by the black hole regularity results of Ref.~\cite{ramos}. In this work, we have indeed found that stellar sensitivities
for the $\a=\b=0$ case can be obtained quite easily, without having to study the stellar interior in detail 
(unlike what had to be done in the general case by Refs.~\cite{Yagi:2013ava,Yagi:2013qpa}). We have found that
for generic matter equations of state,
 the sensitivities vanish exactly when $\a=\b=0$, irrespective of the value of the third coupling constant of the theory, $\l$. This implies in particular that no deviations from general relativity appear in the generation of gravitational waves from neutron star/pulsar binaries at the lowest PN order, i.e. 
no dipole fluxes are present and no deviations from the quadrupole formula appear. Similarly, no deviations from general relativity
arise in the conservative dynamics at Newtonian order. 

We note that this result generalizes the similar result of Ref.~\cite{ramos}, which recently found that {\it black hole} sensitivities vanish exactly
when $\a=\b=0$. In fact, exactly as in the black hole case~\cite{ramos}, we find that the geometry 
of an isolated star moving slowly relative to the preferred
frame matches the predictions of general relativity (at least once asymptotically flat boundary conditions are imposed), and thus
coincides with the geometry of a star at rest relative to the preferred frame.
This fact, obvious in general relativity because of Lorentz symmetry, is non-trivial
in khronometric theory, where Lorentz symmetry is violated. 
We also stress that while the metric shows no deviations from general relativity at first order in
velocity, the \ae ther field is non-trivial and corresponds to a flow moving toward the star at spatial infinity. Necessarily, however, 
the \ae ther stress energy (and thus its backreaction on the metric) vanish everywhere, i.e. the \ae ther behaves as a ``stealth'' field.

Let us also note that our findings
 confirm explicitly, in this particular case, the results of Refs.~\cite{Bellorin:2010je,2014arXiv1407.1259L}, which showed, via a   Hamiltonian analysis,
that khronometric theory (in vacuum and with asymptotically flat boundary conditions) should reduce to general relativity exactly for $\a=\b=0$.
We note, however, that the results presented here for the exterior geometry, and namely the exact solution for $h$ [Eq.~\eqref{hsol}], show that
even in vacuum there exist asymptotically flat solutions to khronometric theory with $\a=\b=0$ that do {\it not} match those of general relativity, even
though they generically present (presumably naked) singularities. These vacuum solutions can indeed be obtained by setting
$k_2=-2 k_1/(3 r_s^3)$ in Eq.~\eqref{hsol}.

Finally, let us note that even if one were willing to accept curvature singularities at the universal horizon of moving black holes in 
khronometric theory,  
the dimensionless parameters $\a$ and $\b$ would still have to be very small to pass experimental bounds ($|\alpha|\lesssim 10^{-5}$, $|\beta|\lesssim 10^{-15}$). Therefore, the sensitivities for finite but experimentally viable $\a$ and $\b$ are expected to be of the order
of $|\sigma|\approx|\sigma(\a=\b=0)+(\partial \sigma/\partial \a) \a+(\partial \sigma/\partial \b) \b|\lesssim 10^{-5}$ (assuming the derivatives in this Taylor expansion are of order $\sim 1$). These values seem outside the reach of not only existing gravitational wave detectors, but also of third generation ones/LISA~\cite{Barausse:2016eii,chamberlain}, because in the limit $\a,\b\to 0$ and $\l\neq0$ the spin-0 speed diverges and the khronon becomes non-dynamical, which
further suppresses deviations from general relativity in the gravitational fluxes (c.f. Sec.~\ref{GW}). 
However, sensitivities $\sigma \sim10^{-5}$ may be testable with observations of the Newtonian dynamics of  triple pulsar systems (for instance
PSR~J0337+1715~\cite{Archibald:2018oxs}). Tests of the 1PN dynamics of binary pulsars
may also yield useful constraints, though they require computing not only the sensitivities but also
their derivatives $\sigma'$. Moreover, further constraints on khronometric gravity might come from cosmological observations, which might have the potential to further constrain $\l$, whose viable range 
 ($|\lambda|\lesssim 0.01-0.1$) is still sizable.

\begin{acknowledgments}
We thank D. Blas, C. Will, K. Yagi and N. Yunes
for insightful conversations on various aspects of the research presented here.
We acknowledge financial support provided under the European Union's H2020 ERC Consolidator Grant
``GRavity from Astrophysical to Microscopic Scales'' grant agreement no. GRAMS-815673.
 This work has also been supported by the European Union's Horizon 2020 research and innovation program under the
Marie Sklodowska-Curie grant agreement No 690904.
\end{acknowledgments}

\bibliographystyle{apsrev4-1}

\bibliography{shortbib}
\end{document}